\documentclass[aps,prl,twocolumn,groupedaddress]{revtex4}
\usepackage{latexsym}
\usepackage{dcolumn}
\usepackage[dvips]{graphicx}
\usepackage{amssymb}
\usepackage{graphics}
\usepackage{amsmath}
\usepackage{epsf}

\begin{document}
\title{Insulating behavior in metallic
  bilayer graphene: Interplay between density inhomogeneity and temperature}
\author{E. H. Hwang and S. Das Sarma} 
\affiliation{Condensed Matter Theory Center, Department of 
        Physics, University of Maryland, College Park, MD 20742}
\begin{abstract}
We investigate bilayer graphene transport in the presence of
electron-hole puddles induced by long-range charged impurities in the
environment.
We explain the insulating behavior observed in the temperature
dependent conductivity of low mobility bilayer graphene using an
analytic statistical theory taking into account the non-mean-field
nature of transport in the highly inhomogeneous density and potential
landscape.  We find that the puddles can induce, even far from the
charge neutrality point, a coexisting metallic and insulating
transport behavior due to the random local activation gap in the
system.  
\end{abstract}
\maketitle


Recent experiments
\cite{zhu2009,feldman2009,fuhrer2010,junzhu2010,herrero2010,ki2010} 
have revealed an intriguingly  strong (and 
anomalous) ``insulating'' temperature dependence in the measured
electrical conductivity of bilayer graphene (BLG) samples, not only at
the charge neutrality point (CNP) where the electron-hole bands touch
each other (with vanishing average carrier density), but also at
carrier densities as high as $10^{12}$ cm$^{-2}$ or
higher. (``Insulating'' temperature dependence of conductivity
$\sigma(T)$ simply means an increasing $\sigma$ with increasing
temperature at a fixed gate voltage, which is, in general, considered
unusual in a nominally metallic system where the resistivity, not the
conductivity, should increase with temperature.) Such an anomalous
insulating temperature dependence of $\sigma(T)$ is typically not
observed in monolayer graphene (MLG) away from the CNP
although the gate
voltage (or equivalently, the density) dependence of MLG and BLG
conductivities are very similar with both manifesting
linear-in-density conductivity away from the CNP and an approximately
a constant minimum conductivity around the CNP \cite{morozov2008,xiao2009}. 

In this Letter we
theoretically establish that this anomalous insulating BLG $\sigma(T)$
behavior is likely to be caused by the much stronger BLG density
inhomogeneity \cite{dassarma2010} (compared with MLG) which gives rise
to a qualitatively 
new type of temperature dependence in graphene transport,
namely, the intriguing coexistence of both metallic and activated transport,
hitherto not
discussed in the literature. 
We therefore predict that the observed temperature dependence of BLG
$\sigma(T)$ arises from the same charged impurity induced puddles in
the system which are responsible for the minimum conductivity plateau
at the CNP \cite{dassarma2010b}. We provide an analytic theory which
appears to be in 
excellent qualitative agreement with the existing experimental
results. One direct prediction of our theory, the suppression of the
anomalous insulating temperature dependence in high mobility samples
with lower disorder, seems to be consistent with experimental
observations.
As a direct corollary of our theory, we find, consistent with
experimental observation \cite{castro2007,oostinga2008,mak2009}, that a gapped
BLG (with the gap 
at the CNP induced, for example, by an external electric field) would
typically manifest a transport activation gap substantially smaller
than the intrinsic spectral gap (i.e. the energy band gap) unless the
band gap is much larger than the typical puddle-induced potential
energy fluctuations. 

Our theory is based on a physically motivated idea: In the presence of
large potential fluctuations $V({\bf r})$, the local Fermi level,
$\mu({\bf r}) = E_F - V({\bf r})$, would necessarily have large spatial
fluctuations [particularly when $E_F \alt s$, where $s=V_{rms}$ is the
root-mean-square fluctuations or the standard deviation in $V({\bf r})$],
leading to a complex temperature dependence of transport since both
metallic and activated transport would be present due to random local gap.
Below we carry out an analytical theory implementing this
physical idea. We will see that this physical idea leads to the
possible coexistence of metallic and activated transport, which
explains the observed temperature dependence of BLG transport.

We start by assuming that the disorder-induced potential energy
fluctuations in the BLG is described by a distribution function $P(V)$
which $V=V({\bf r})$ is the fluctuating potential energy at the
point ${\bf r}\equiv (x,y)$ in the 2D BLG plane. 
We approximate the probability $P(V)dV$ of finding the
local electronic potential 
energy within a range $dV$ about $V$ to be a Gaussian form, i.e.,
$P(V) = \frac{1}{\sqrt{2\pi s^2}} \exp(-V^2/2s^2)$,
where $s$ is the standard deviation (or equivalently, the strength of
the potential fluctuation).
Then in the presence of electron-hole puddles
the density of states (DOS) is reduced by the allowed electron region
fraction and given by
$D(E) =  \int_{-\infty}^{E}D_0P(V)dV 
     =  {D_0}{\rm erfc}(-E/\sqrt{2}s)/2$,
where erfc$(x)$ is the complementary error function and 
$D_0 = {g_sg_v m}/(2\pi \hbar^2)$ is
the DOS in a homogeneous system,
where $m$ is the band effective mass, $g_s=2$ and $g_v=2$ are the spin and
valley degeneracies, respectively.
We have $D_0=2.8\times 10^{10}$ cm$^{-2}$/meV with the effective mass
$m=0.033m_e$ (where $m_e$ is the bare electron mass).
Note that the tail of the DOS is determined by the potential fluctuation
strength $s$.

\begin{figure}
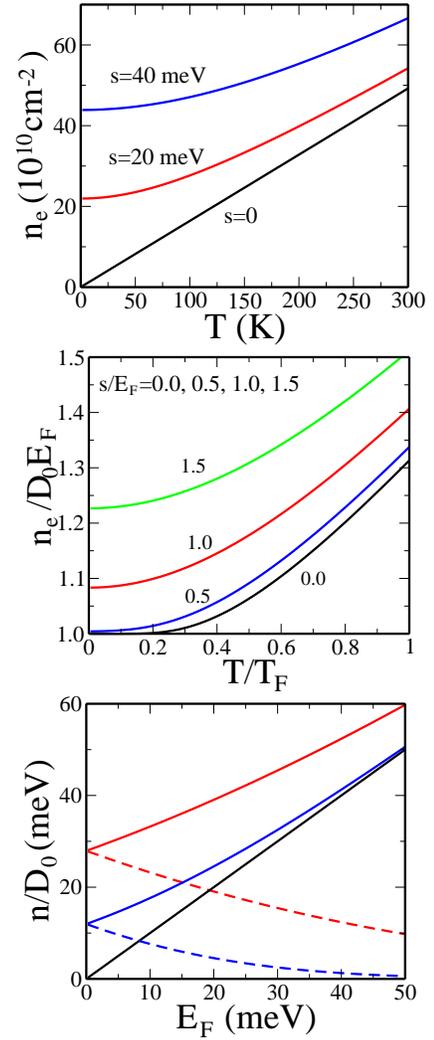

\epsfysize=1.8in
\epsffile{fig_1a.eps}
\epsfysize=1.8in
\epsffile{fig_1b.eps}
\epsfysize=1.8in
\epsffile{fig_1c.eps}
\caption{ (Color online)
(a) The electron density at CNP as a function of
  temperature for different   $s$. At $T=0$ the density is given by
  $n_0=D_0 s/\sqrt{2\pi}$. 
(b) The temperature dependent electron density at finite $E_F$
for different $s$. For $s/E_F \neq 0 $
  the leading order behavior is quadratic while at $s=0$ the density
  is exponentially suppressed.
(c) Total electron densities (solid lines) and hole densities (dashed
lines) as a function of $E_F$ for two
different $s=30$ meV and 70 meV. The linear line represents the density
difference $n=n_e-n_h=D_0E_F$, which linearly depends on the Fermi
energy. The densities at the band tails are given by $n_e(E_F=0)=n_h(E_F=0)=D_0
s/\sqrt{2\pi}$.
\label{fig:den}
}
\end{figure}

Since BLG is a gapless semiconductor the
electron density at
finite temperature increases due to the direct thermal excitation from
valence band to conduction band, and this thermal excitation is
an important source of temperature dependent transport. 
Thus, we first consider the temperature dependence of thermally
excited electron density.
The total electron density is given by
\begin{equation}
n_e  = \int_{-\infty}^{\infty}D(E)\frac{dE}{e^{\beta(E-E_F)}+1},
\end{equation}
where $\beta=1/k_BT$ and $E_F$ is the Fermi energy.
When the Fermi energy is zero (or at CNP) all
electrons are located in the band tail at $T=0$ and 
the electron density in the band tail  is given by
$n_0=n_e(E_F=0)= {D_0 s}/{\sqrt{2\pi}}$.
Note that the electron density in the band tail is linearly
proportional to the standard deviation $s$.
At finite temperatures we find the asymptotic behavior of $n_0(T)$. 
The low temperature ($k_BT/s \ll 1$) behavior of electron density at
CNP becomes 
\begin{equation}
n_e(T) = n_0 \left [ 1 + \frac{\pi^2}{6} \left ( \frac{k_BT}{s} \right
  )^2 \right ].
\label{eq:den_0}
\end{equation}
Thus, the electron density increases quadratically in low temperature
limit. For homogeneous BLG with the constant DOS
the electron density at finite temperatures is
given by
$n_e(T)=D_0\ln(2)k_BT$.
The presence of the band tail suppresses the thermal excitation of
electrons and gives rise to the quadratic behavior. However,
at high temperature the density increases linearly with the same
slope as in the homogeneous system, i.e.,
\begin{equation}
n(T) \sim D_0 \left [ \ln(2) k_BT + \frac{1}{8}\frac{s^2}{(k_BT)^2}
  \right ].
\label{eq:den_0h}
\end{equation}
In Fig.~\ref{fig:den}(a) we show the temperature dependent electron
density at CNP for different standard deviations.

In the case of finite doping (or gate voltage), i.e., the Fermi level
away from CNP, $E_F\neq 0$, the
electron density of the homogeneous BLG for $s=0$ is given by 
\begin{equation}
n_{0e}(T)=D_0E_F \left [1+ t \ln \left (1+e^{-1/t}
  \right ) \right ],
\end{equation}
where $t=T/T_F$ and $T_F = E_F/k_B$. At low temperatures ($T \ll T_F$)
the thermal excitation 
is exponentially suppressed due to the Fermi function, but
at high temperatures ($T \gg T_F$) it increases linearly. 
In the presence of electron-hole puddles ($s \neq 0$) we have the
electron density at zero temperature for the inhomogeneous system:
\begin{equation}
n_e(0) = {D_0 E_F} \left [ \frac{1}{2}{\rm erfc} \left(
  \frac{-1}{\sqrt{2}\tilde{s}} 
  \right ) + \frac{\tilde{s}}{\sqrt{2\pi}} e^{-1/2\tilde{s}^2} \right ],
\end{equation}
where $\tilde{s}=s/E_F$. At low  temperatures ($T \ll T_F$) the
asymptotic  
behavior of the electron density is given by
\begin{equation}
n_e(T) = n_e(0) + D_0 E_F  \frac{\pi^2
}{12\sqrt{2}}\frac{e^{-1/2\tilde{s}^2}}{\tilde{s}} 
\left ( \frac{T}{T_F} \right )^2.
\label{eq:den_mu}
\end{equation}
The leading order term is the same quadratic behavior as in
undoped BLG ($E_F=0$), but the coefficient is strongly
suppressed by fluctuation. In the case of $s > E_F$,
the existence of electron-hole puddles gives rise to a notable
quadratic behavior [see Fig.~\ref{fig:den}(b)]. At high temperatures
  ($T \gg T_F$)  we find 
\begin{equation}
n_e(T)= n_{0e}(T)  +\frac{D_0E_F}{(1+e^{\beta E_F})^2}
\frac{\tilde{s}^2}{2} \frac{T_F}{T}. 
\end{equation}

At CNP ($E_F=0$) electrons and holes are
equally occupied. As the Fermi energy increases, more electrons occupy
increasingly larger proportion of space [see
  Fig.~\ref{fig:den}(c)]. For $E_F \gg s$ nearly all 
space is allowed to the electrons, and the conductivity of the system
approaches the characteristic of the homogeneous materials.
In the presence of electron-hole puddles, there is 
a possible coexistence of metallic and thermally-activated
transport.
When electron puddles occupy more space than hole puddles, most
electrons follow the
continuous metallic paths extended throughout the system, but it is
possible at finite temperature that
the thermally activated transport of electrons persists above the hole puddles.
On the other hand, holes in hole puddles propagate freely, but when they
meet electron puddles activated holes conduct over the electron puddles.
Carrier transport in each puddle is characterized by propagation of
weak scattering transport theory \cite{dassarma2010}.
The activated carrier transport of prohibited regions, where
the local potential energy is $V$ less (greater) than Fermi energy for
electrons (holes),
is proportional to the Fermi factor. If $\sigma_e$ and $\sigma_h$ are
the average conductivity of electron and hole puddles,
respectively, then
the activated conductivities are given by 
\begin{subequations}
\begin{eqnarray}
\sigma_e^{(a)}(V) & = & \sigma_e \exp[\beta (E_F-V)], \\
\sigma_h^{(a)}(V) & = &\sigma_h \exp[\beta (V-E_F)],
\end{eqnarray}
\end{subequations}
where the density and temperature dependent average conductivities
($\sigma_e$ and $\sigma_h$) are
given within the Boltzmann transport theory \cite{dassarma2010} by 
$\sigma_{e} = {n_e e^2 \langle \tau \rangle}/{m}$ and 
$\sigma_{h} = {n_h e^2 \langle \tau \rangle}/{m}$,
where $n_e$ and $n_h$ are average electron and hole densities,
respectively, and $\langle \tau \rangle$ is the transport relaxation
time which depends explicitly on the scattering mechanism
\cite{dassarma2010}. 

Now we denote the electron (hole) puddle as region `1' (`2'). 
In region 1 electrons are occupied more space than holes when $E_F>0$.
The fraction of the total area occupied by electrons with
Fermi energy $E_F$ is given by
$p=\int_{-\infty}^{E_F}P(V)dV$.
Then the
total conductivity of region 1 can be calculated
\begin{eqnarray}
\sigma_1 & = & \frac{1}{p}\int^{E_F}_{-\infty}(\sigma_e +
\sigma_h^{(a)})P(V) dV, \nonumber \\
&=& \sigma_{e}+\frac{\sigma_{h}}{2p} e^{
    \frac{\beta^2s^2}{2} -\beta E_F } {\rm erfc} \left (
    -\frac{E_F}{\sqrt{2} s} + \frac{\beta s}{\sqrt{2}} \right).
\label{eq:sig1}
\end{eqnarray}
At the same time the holes occupy the area with a fraction $q=1-p$ and
the total conductivity of region 2 becomes
\begin{eqnarray}
\sigma_2 & = &\frac{1}{q}\int_{E_F}^{\infty}(\sigma_e^{(a)} + \sigma_h)P(V) dV
\nonumber \\
&=& \sigma_{h}+\frac{\sigma_{e}}{2q} e^{
    \frac{\beta^2s^2}{2} +\beta E_F } {\rm erfc} \left (
    \frac{E_F}{\sqrt{2} s} + \frac{\beta s}{\sqrt{2}} \right).
\label{eq:sig2}
\end{eqnarray}
The $\sigma_1$ and $\sigma_2$ are distributed according to the
binary distribution.
The conductivity of binary system can be calculated by using
the effective medium theory of conductance in
mixtures\cite{kirkpatrick1973}. The result 
for a 2D binary mixture of components with conductivity $\sigma_1$ and
$\sigma_2$ is given by \cite{kirkpatrick1973}
\begin{equation}
\sigma_t = (p-\frac{1}{2})\left [ (\sigma_1 -\sigma_2) +
  \sqrt{(\sigma_1-\sigma_2)^2+\frac{4\sigma_1 \sigma_2}{(2p-1)^2}} 
  \right ].
\label{eq:sig_tot}
\end{equation}
This result can be applied for all Fermi energy. For a large doping
case, in which the hole puddles disappear, we have $p=1$ and
$\sigma_2=0$, then Eq.~(\ref{eq:sig_tot}) becomes $\sigma = \sigma_1$,
i.e., the conductivity of electrons in the homogeneous system.

\begin{figure}
\includegraphics[width=6.5cm]{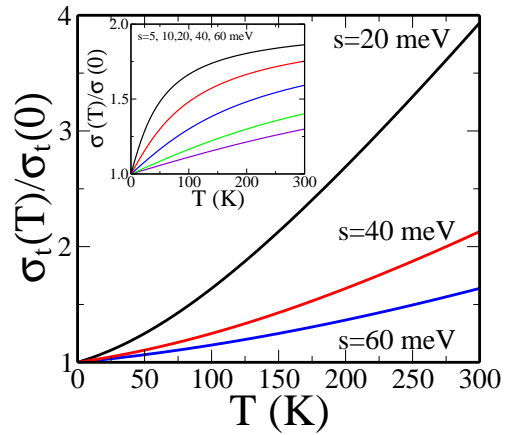}
\caption{(Color online) 
$\sigma_t(T)$ at charge neutral point for
different $s$. Inset shows the thermally activated conductivity as a
function temperature. 
\label{fig:sig_mu0}
}
\end{figure}

\begin{figure}[tb]
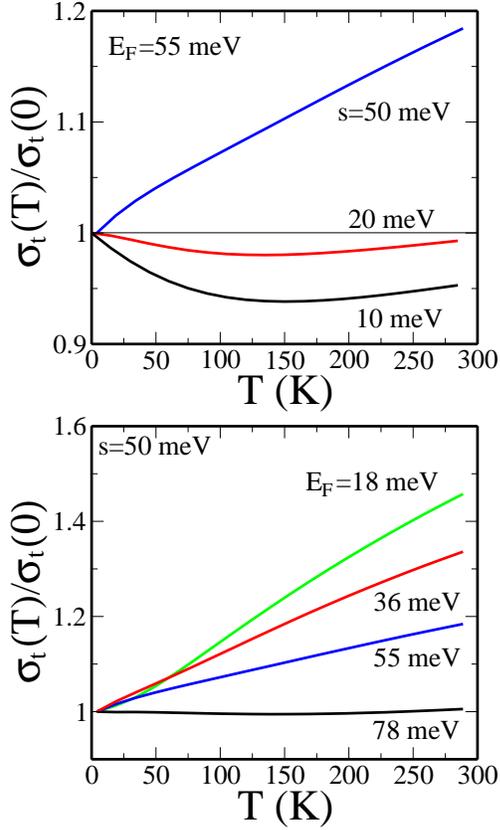

 \begin{center}
  \includegraphics[width=6.5cm]{fig_3a.eps} 
  \includegraphics[width=6.5cm]{fig_3b.eps} 
  \caption{ 
           (Color online). 
           (a) $\sigma_t(T)$ for  $E_F=55$ meV and for
    different $s$. 
           (b) $\sigma_t(T)$
for $s=50$ meV and for
    several $E_F=18$, 36, 55, 78 meV, which correspond to the
    densities $n=0.5,$ 1.0, 1.5, 2.0$\times 10^{12}$ cm$^{-2}$.   
          } 
  \label{fig:sig_tot}
 \end{center}
\end{figure} 

We first consider the conductivity at CNP
($E_F=0$). The conductivities in each region are given by
\begin{subequations}
\begin{eqnarray}
\sigma_1 & = &  \sigma_{e} \left [ 1 + \frac{\eta}{2p} e^{\beta^2 s^2/2}
  {\rm erfc} (\beta s/\sqrt{2}) \right ], \\
\sigma_2  & = & \sigma_{h} \left [ 1 + \frac{1}{2q\eta} e^{\beta^2 s^2/2}
  {\rm erfc} (\beta s/\sqrt{2}) \right ],
\end{eqnarray}
\end{subequations}
where $\eta = n_h/n_e$ is the ratio of the hole density to the
electron density. 
Since the electrons and holes are equally populated we have $p=q=1/2$
and $\sigma_{e}=\sigma_{h}$, then the total conductivity becomes
$\sigma_{t}  = \sqrt{\sigma_1 \sigma_2} = \sigma_1$.
The asymptotic behavior of the conductivity at low temperatures ($k_BT
\ll s$) becomes 
\begin{equation}
\sigma_t(T) = \sigma_{e} \left [1 + \sqrt{ \frac{2}{\pi}} \frac{k_BT}{s}
  - \frac{2}{\sqrt{\pi}}\frac{(k_BT)^3}{s^3} \right ].
\end{equation}
The activated conductivity increases linearly with a slope
$\sqrt{2/\pi}k_B/s$ as temperature increases. 
Because 
$s$ is typically smaller in higher mobility sample, 
the high mobility samples show stronger insulating behavior at low
temperatures. The next order temperature correction to  conductivity
arises from the 
thermal excitation given in Eq.~(\ref{eq:den_0}) which gives $T^2$
corrections. Thus in low temperature limit the total conductivity at 
CNP is given by
\begin{equation}
\sigma_t(T) = \sigma(0) \left [1+ \sqrt{\frac{2}{\pi}}\frac{k_BT}{s} +
  \frac{\pi^2}{6} \left ( \frac{k_BT}{s} \right )^2 \right ].
\end{equation}
At high temperatures ($k_BT \gg s$) we have
\begin{equation}
\sigma_t = \sigma_e \left [ 2 - \sqrt{ \frac{2}{\pi}} \frac{s} {k_BT} +
  \frac{s^2}{2 (k_BT)^2} \right ].
\end{equation}
The total conductivity due to the activation behavior approaches
a limiting value and all temperature dependence comes from the thermal
excitation through the change of carrier density given in
Eq.~(\ref{eq:den_0h}). Thus at very high temperatures ($T\gg s/k_B$)
the BLG conductivity at the charge neutral point increases linearly
with a universal 
slope $\ln(2)$ regardless of 
the sample quality. In Fig.~\ref{fig:sig_mu0} we show the calcuated
temperature dependent conductivity at charge neutral point.

At finite doping ($E_F > 0$) the temperature dependent conductivities
are very complex because three energies ($E_F$, $s$, and
$k_BT$) are competing. Especially when $k_BT \ll s$, regardless of
$E_F$, we have the asymptotic behavior of
conductivities in region 1 
and 2 from Eqs.~(\ref{eq:sig1}) and (\ref{eq:sig2}), respectively,
\begin{subequations}
\begin{eqnarray}
\sigma_1 & = & \sigma_{e} \left [ 1+ \frac{\eta}{2p} e^{-1/2\tilde{s}^2}
\sqrt{\frac{2}{\pi}}  \frac{1}{\tilde{s}/t-1/\tilde{s}} \right ], \\
\sigma_2 & = & \sigma_{h} \left [ 1+ \frac{1}{2q \eta} e^{-1/2\tilde{s}^2}
\sqrt{\frac{2}{\pi}}  \frac{1}{\tilde{s}/t+1/\tilde{s}} \right ],
\end{eqnarray}
\end{subequations}
where $\tilde{s}=s/E_F$ and $t=T/T_F$. The leading order correction is
linear but the coefficient is exponentially suppressed by the term
$\exp(-E_F^2/2s^2)$. This fact indicates that in the high mobility sample
with small $s$, the activated conductivity is weakly temperature
dependent except around CNP, i.e. $E_F < s$. 
Since the density increase by thermal excitation is also suppressed
exponentially by the same factor [see Eq.~(\ref{eq:den_mu})] the
dominant temperature dependent conductivity arises from the scattering
time \cite{dassarma2010}.
On the other hand,
for a low mobility sample with a large $s$, the linear temperature
dependence due to thermal activation can be observed even at high
densities $E_F \agt s$.

In Fig.~\ref{fig:sig_tot} we show the total conductivities
(a) for a fixed $E_F$ and several $s$ and (b)
for a fixed $s$ and several $E_F$.  In total conductivity the activated
insulating behavior competes with the metallic behavior due to the
temperature dependent screening effect. When $s$ is small the
activated behavior is suppressed. As a result the total conductivity
manifests the metallic behavior \cite{dassarma2010}. However, for large $s$ the
activated temperature dependence 
behavior overwhelms the metallic temperature dependence,
and the system shows 
insulating behavior.

Finally, we discuss three important issues: (1) The same physics, of
course, also applies to MLG graphene, but the quantitative effects of
inhomogeneity (i.e. the puddles) are much weaker since simple
estimates show that the dimensionless potential fluctuation strength
$\tilde s$ ($\equiv s/E_F$) is much weaker in MLG than in BLG because
of the linear (MLG) versus constant (BLG) DOS in the two systems. In
particular, $\tilde s_{BLG}/ \tilde s_{MLG} \sim 32 \sqrt{\tilde n}$
where $\tilde n = n/10^{10}$, and therefore $\tilde s_{BLG} \gg \tilde
s_{MLG}$ upto $n=10^{13}$ cm$^{-2}$.
Direct calculations \cite{dassarma2010} show that the self-consistent
values of $s$ tend to be much larger in BLG than in MLG for identical
impurity disorder. 
In very low mobility MLG samples,
where $s$ is very large, 
the insulating behavior of temperature dependent resistivity  can be
observed at high 
densities even in MLG samples\cite{tan2007,heo2010}. 
(2) We have neglected all quantum 
tunneling effects in our consideration because they are unimportant
except at very low temperatures. In particular, Klein tunneling is
strongly suppressed in strong disorder \cite{rossi2010}. (3) In the
presence of a BLG gap ($\Delta_g$), the situation becomes extremely
complicated since four distinct energy scales ($s$, $E_F$, $k_BT$,
$\Delta_g$) compete, and any conceivable temperature dependence
may arise depending on the relative values of these four energy
scales. It is, however, obvious that any experimental measurement of
the activation gap ($\Delta_a$) in such an inhomogeneous situation
will produce $\Delta_a \ll \Delta_g$ unless $\Delta_g \gg s$. The
system is now dominated by a random local gap arising from the
competition among $s$, $\Delta_g$, and $E_F$, and no simple activation
picture would apply. This is precisely the experimental observations
\cite{castro2007,oostinga2008,mak2009}.

Work supported by ONR-MURI and NRI-NSF-SWAN.



\end{document}